\documentclass[twocolumn,prl]{revtex4}

\usepackage{amsmath,amssymb}
\usepackage{graphicx}
\usepackage{dcolumn}
\usepackage{bm}

\begin{document}


\title{Interplay of Peltier and Seebeck effects in nanoscale nonlocal spin valves}

\author{F. L. Bakker}
\email{F.L.Bakker@rug.nl}
\author{A. Slachter}
\author{J.-P. Adam}
\author{B. J. van Wees}

\affiliation{Physics of Nanodevices, Zernike Institute for Advanced Materials, University
of Groningen, The Netherlands
}%

\date{\today}

\begin{abstract}

We have experimentally studied the role of thermoelectric effects in nanoscale nonlocal spin valve devices. A finite element thermoelectric model is developed to calculate the generated Seebeck voltages due to Peltier and Joule heating in the devices. By measuring the first, second and third harmonic voltage response non locally, the model is experimentally examined. The results indicate that the combination of Peltier and Seebeck effects contributes significantly to the nonlocal baseline resistance. Moreover, we found that the second and third harmonic response signals can be attributed to Joule heating and temperature dependencies of both Seebeck coefficient and resistivity.

\end{abstract}

\maketitle

The Seebeck and the related Peltier effects are the fundamental phenomena of thermo electricity, a field subject to extensive research during the previous decades~\cite{Barnard1972}. Despite the fact that these are bulk material properties, they can be utilized to measure the temperature or to generate heat locally at or close to an interface between different materials. While progress in nanoscale device fabrication has made it possible to study these phenomena at continuously decreasing length scales, they are rarely taken into account to analyze electrical measurements in nanostructures. In the specific field of spintronics, a detailed understanding of the interaction between heat transport and the charge and spin degrees of freedom is highly required \cite{valenzuela2009}. This emerging branch, called \emph{spin caloritronics} \cite{Bauer2010}, has recently drawn considerable attention \cite{Dubi2009,Slachter2010,Gravier2006,haiming2010,Uchida2008}, and (spin-) thermoelectric effects have been experimentally examined in magnetic multilayer nanostructures \cite{Gravier2006,haiming2010} and in macroscopically large ferromagnetic strips \cite{Uchida2008}. In this Letter, we use lateral nonlocal spin valve devices as a tool to study the interplay between heat, charge and spin at the nanoscale. The nonlocal device design enables us to separate the charge and heat current, and hence, excludes spurious effects. We find that the baseline resistance in nonlocal spin valve measurements originates mainly from Peltier heating/cooling at the injector junction and the Seebeck effect at the detector junction. Furthermore, we demonstrate that it is experimentally feasible to use basic thermoelectrics to obtain control over the heat flow in nanostructures.

\begin{figure}[b]
\includegraphics[width=8.5cm]{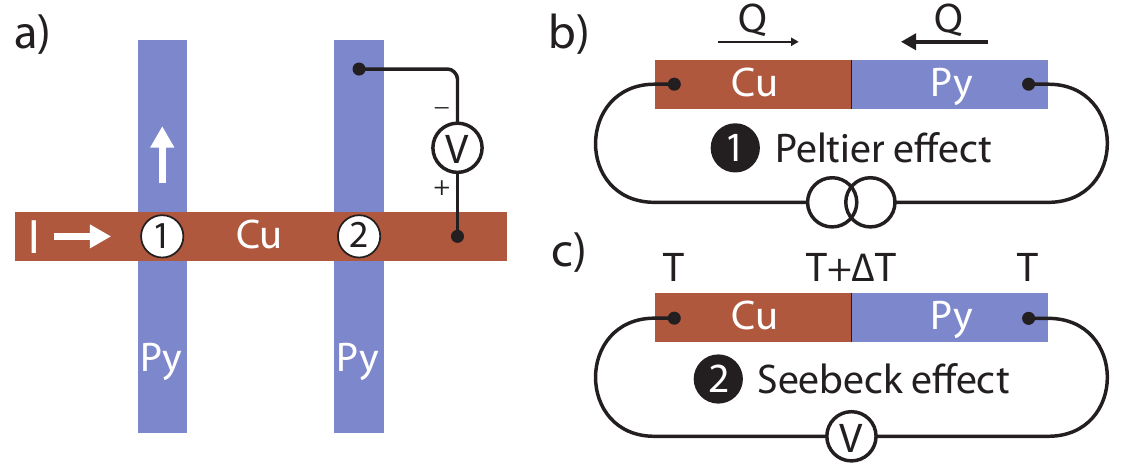}
\caption{\label{fig:concept} (color online) a) Schematic drawing of a typical spin valve experiment. Current is sent through the first Cu/Py interface, while the voltage drop is measured at the second interface. b) Due to the difference in the Peltier coefficients for Cu and Py, the heat current $Q$ carried by the electrons changes across the interface. Hence, interface 1 is locally heated or cooled depending on the direction of the current. c) The second interface acts as a thermocouple and detects the local electron temperature via the Seebeck effect.}
\end{figure}

The nonlocal spin valve experiment is schematically depicted in Fig.\ \ref{fig:concept}a. Two permalloy (Py) electrodes are overlapped with a Cu strip, creating two ferromagnetic/nonmagnetic metal (F/N) interfaces. Spin injection across a F/N interface is well-described in terms of a two current model \cite{ValetFert1993}. When a spin-polarized current flows across a F/N interface, the sudden change in spin-dependent conductivity causes a spin splitting of the electrochemical potential \cite{vanSon1987}. As a result, a spin accumulation builds up in the region close to the interface and decays with the spin relaxation length. Electrical spin injection and detection in a nonmagnetic metal was demonstrated first by Johnson and Silsbee \cite{Johnson1985} and succeeded later by Jedema \emph{et al.}\ \cite{Jedema_N_410_2001} in a lateral structure, at room temperature, by performing nonlocal spin valve measurements. Here, a spin current is injected into the Cu strip by sending a charge current through the first F/N interface (Fig.\ \ref{fig:concept}a). A spin voltage can be detected at the second interface provided that the spacing between injector and detector is shorter than the spin relaxation length of the Cu \cite{Jedema_N_410_2001}.

Ideally, the voltage detected at the second interface in the nonlocal geometry will be zero in the absence of a spin accumulation. Since the current and voltage path are completely separated, one expects no Ohmic voltage drop on the right side of interface 1 (Fig.\ \ref{fig:concept}a). The voltage arising from a spin accumulation is then bipolar, having equal magnitude but opposite sign for the parallel and antiparallel alignment of both ferromagnets. However, the baseline resistance observed in experiments is in general nonzero \cite{Jedema_N_410_2001}. Current spreading at the injector can account for an Ohmic resistance at the detector, as discussed by Johnson and Silsbee \cite{Johnson2007}. The obtained voltage $V_r$ that results from this is found to depend exponentially on the separation $L$ between the two interfaces as $V_r \propto e^{-\pi L/W}$, with $W$ the width of the Cu strip. Moreover, spin dependent scattering at the detector interface \cite{Garzon2005} has been invoked to explain an offset voltage, but both effects are not sufficient to describe the data accurately. 

Here we address a new origin of the baseline resistance, composed of thermoelectric phenomena which are generally disregarded in the analysis. We show that the Peltier and the related Seebeck effect, when combined in a lateral nanostructure, give rise to a significant modification of the baseline resistance (Fig.\ \ref{fig:concept}). 

If an electrical current $I$ flows through a Cu/Py interface, due to the mismatch of the Peltier coefficients at the interface, heat accumulates or is absorbed at the interface. The heat current $Q$ carried by the electrons, represented by the Peltier coefficient, is different on both sides of the interface. Since the charge current is continuous across the interface, the heat current has a discontinuity. Consequently, the interface is heated or cooled depending on the sign of the current (Fig.\ \ref{fig:concept}b). The inverse process, called the Seebeck effect, refers to the generation of a voltage $V$ by a temperature gradient $\nabla T$. This effect can be exploited to probe the local electron temperature at or close to the interface, similar to the functioning of a thermocouple (Fig.\ \ref{fig:concept}c). As copper is an excellent thermal conductor, the heat generated at interface 1 can be efficiently transferred to interface 2 and is then, via the Seebeck effect, translated back into a voltage.

In order to quantify the Peltier and Seebeck effects a thermoelectric model is developed. The charge current density $J$ and the heat current density $Q$ in these nanostructures can be related to the voltage and temperature in the following way:
\begin{equation}
\begin{pmatrix} \vec{J} \\ \vec{Q} \end{pmatrix} = \begin{pmatrix} -\sigma & \sigma S \\ \sigma\Pi & -k  \end{pmatrix} \begin{pmatrix} \vec{\nabla} V \\ \vec{\nabla} T \end{pmatrix}
\end{equation}
with $\sigma$ the electrical conductivity, $k$ the thermal conductivity, $S$ the Seebeck coefficient and $\Pi=ST$ the Peltier coefficient. Joule heating is incorporated via $\nabla Q = J^2/\sigma$. The model can be extended by introducing spin-dependent conductivities $J_{\uparrow,\downarrow}=-\sigma_{\uparrow,\downarrow}/e \nabla \mu_{\uparrow,\downarrow}$, with $\sigma_{\uparrow,\downarrow}$ and $\mu_{\uparrow,\downarrow}$ the spin-dependent conductivity and electrochemical potential, respectively \cite{ValetFert1993,Gravier2006_024419}. Bulk spin relaxation is introduced via the equation $\nabla^2 \left( \mu_{\uparrow}-\mu_{\downarrow} \right) = \left( \mu_{\uparrow}-\mu_{\downarrow} \right) / \lambda$, with $\lambda$ the spin relaxation length. 

\begin{figure}[b]
\includegraphics[width=8.5cm]{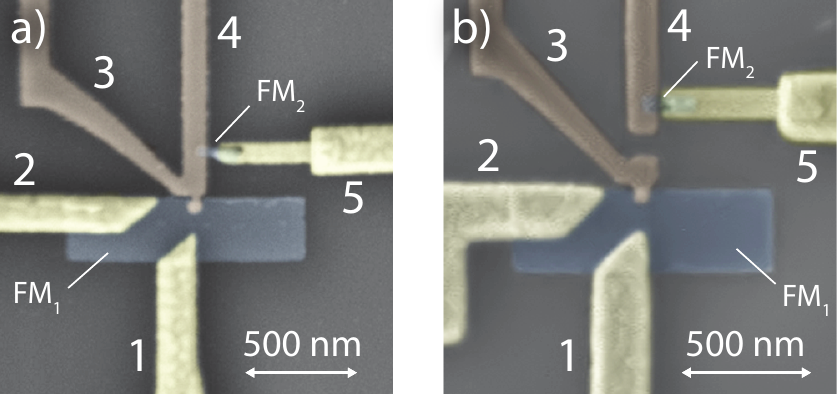}
\caption{\label{fig:sem} (color online) Scanning Electron Microscope (SEM) image of the device lay-out. a) Standard nonlocal spin valve geometry. Current is sent from contact 1 to 3, while the voltage is measured between 4 and 5. Contact 2 is not used. b) Similar device geometry with an electrically isolated detector circuit. }
\end{figure}

Two batches of lateral nonlocal spin valve devices were fabricated on a thermally oxidized Si substrate in a five-step electron beam lithography process. Fig.\ \ref{fig:sem} shows the scanning electron microscope (SEM) images of the two types of devices. A device consists of two 15~nm thick Permalloy (Py) islands, a large injector FM$_1$ (1 $\mu$m $\times$ 300~nm) and a small detector FM$_2$ (150~nm $\times$ 50~nm), separated from each other by a distance $L$. Both ferromagnets are contacted on one side with Au electrodes and with a 60~nm thick Cu strip on the other side. To reduce Joule heating in the leads, the Au contacts on FM$_1$ have a thickness of 170~nm. Furthermore, the Py/Cu interface area of the injector and detector are kept very small (45 $\times$ 50~nm) in order to increase the spin injection and detection efficiency. The metallic layers are deposited using an electron beam evaporator with a base pressure of $1 \times 10^{-6}$~mbar. Prior to deposition of Au and Cu the interfaces are cleaned by Ar ion milling to assure good Ohmic interfaces.

We use a lock-in amplifier for detecting the voltage $V$ across the Cu/FM$_2$ interface, between contact 4 and 5 (Fig. \ref{fig:sem}). Simultaneously an AC current $I$ is sent from contact 1 to 3. If the response of the system is nonlinear, the higher order terms can be extracted separately by measuring the higher harmonics:
\begin{equation}
V = R_1 I + R_2 I^2 + ...
\end{equation}
with $R_i$ ($i=1,2,...$) the i-th harmonic 'resistance' response. The current is applied at a frequency below 1~kHz, much lower than the relevant time scales for thermal conduction in these nanostructures. All electrical measurements are performed at room temperature.

\begin{figure}[t]
\includegraphics[width=8.5cm]{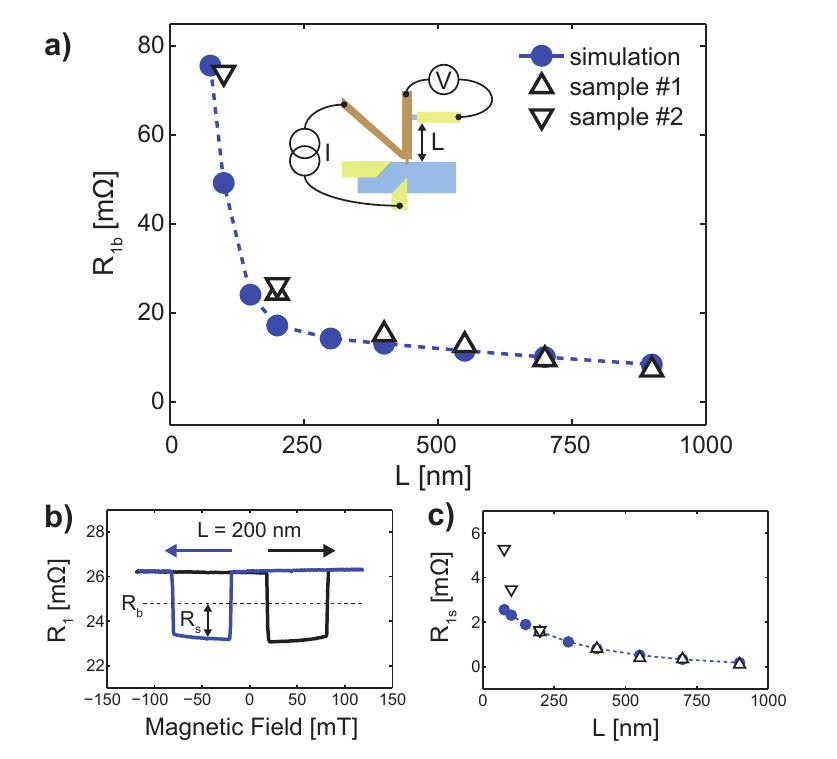}
\caption{\label{fig:data1} (color) a) R$_{1b}$ measured as a function of the spacing $L$ between the two ferromagnets, as indicated by the measurement geometry in the inset. The triangles reflect measurements taken for two different samples, whereas the blue dots correspond to the FEM calculations. b) Nonlocal spin valve measurement with the magnetic field swept back and forth, being indicated by the arrows. $R_{1s}$ is defined as the resistance due to the presence of a spin accumulation and $R_{1b}$ is the baseline resistance. c) $R_{1s}$ as a function of the separation $L$ between both ferromagnets.}
\end{figure}

The first harmonic response $R_1$, reflects the sum of the baseline resistance and a resistance due to the presence of a spin accumulation. The magnetic field dependence of $R_1$ is shown in Fig.\ \ref{fig:data1}b, with $R_{1s}$ the spin valve signal and $R_{1b}$ defined as the baseline resistance. To examine the distance dependence as proposed by Johnson and Silsbee \cite{Johnson2007}, $R_1$ is measured for device type 1 (Fig.\ \ref{fig:sem}a) with different $L$, varying between 75 and 900~nm. The baseline resistance is plotted in Fig.\ \ref{fig:data1}a. If we neglect thermoelectric effects, the baseline is expected to decrease exponentially as $R_{1b} \propto e^{-\pi L/W}$, with $W$ the width of the Cu contact. The data clearly shows an exponential decrease of resistance for small $L$, but decays more slowly for larger separations. Note that the spin valve voltage shows only an exponential dependence due to spin relaxation in the Cu (shown in Fig.\ \ref{fig:data1}c). Assuming that $\lambda_F=\text{5 nm}$ for Py, we deduce from a two current model \cite{ValetFert1993} the spin relaxation length of Cu and bulk current polarization of Py to be 350~nm and 25\%, respectively.

\begin{figure}[t]
\includegraphics[width=8.5cm]{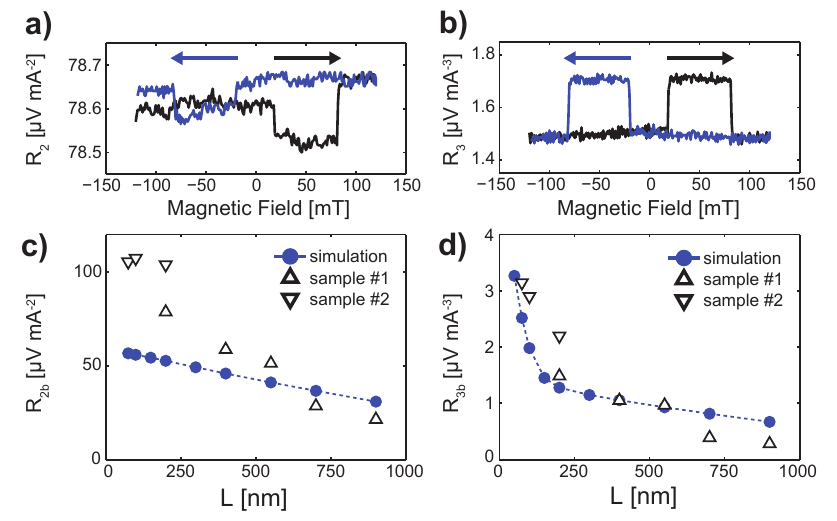}
\caption{\label{fig:data2} (color) a) R$_2$ as a function of magnetic field. The baseline resistance is mainly caused by the Seebeck voltage induced by Joule heating. b) R$_3$ plotted versus magnetic field. The baseline reflects the modification in Seebeck coefficient and resistance due to temperature changes. Likewise, the spin signal indicates how the spin valve signal is altered by temperature. c) Baseline resistance $R_{2b}$ as a function of $L$. Triangles represent data for two different samples, blue dots are simulation results. d) R$_{3b}$ measured versus $L$. The shape of $R_{3b}$ is similar to $R_{1b}$ because it describes the temperature dependence of the effects that generate $R_{1b}$.}
\end{figure}

For the same set of samples, the magnetic field dependence of the higher harmonic responses $R_2$ and $R_3$ is investigated for $L = \text{200~nm}$ (shown in Fig.\ \ref{fig:data2}a and \ref{fig:data2}b). In addition to a nonzero baseline, we observe a spin voltage in $R_2$ and $R_3$ as well. Note that the spin signal has opposite sign for R$_3$ as compared to $R_2$ and R$_1$. The baseline is measured as a function of the separation $L$ between the two ferromagnets and the result is shown in Fig.\ \ref{fig:data2}c and \ref{fig:data2}d. We find that for $R_{2b}$ the exponential relation to $L$ is absent, whereas $R_{3b}$ shows similar behavior as $R_{1b}$, decreasing exponentially for short $L$ and having a much weaker decay for larger separations.

In the following, we show that the observed baselines of $R_{1b}$, $R_{2b}$ and $R_{3b}$ can be attributed to the Peltier and Seebeck effect. The voltage at the detector can be written as the sum of the spin voltage $V_{s}$, a resistive part $V_{r}$ and, as a new ingredient, a Seebeck voltage $S\Delta T$.
$S$ is the effective Seebeck coefficient and $\Delta T$ the local temperature difference close to the Py/Cu interface of FM$_2$. For these devices two important sources of heat exist, Peltier heating at the injector interface and Joule heating in the entire current path. Heat is carried away by thermal transport through the metallic leads and via the SiO$_2$ substrate and consequently, a temperature gradient evolves in the vicinity of FM$_2$. The generated Seebeck voltage is proportional to a combination of the Seebeck coefficients of Py, Cu and Au, the materials that contain a temperature gradient in the detector circuit. This circuit can essentially be seen as a thermocouple with an effective Seebeck coefficient $S$. 

In contrast to Peltier heating, being linear with $I$, Joule heating scales as $\Delta T \propto I^2$. Hence, the thermoelectric contribution to the baseline, $S\Delta T$, originates from Peltier heating for R$_{1}$ and from Joule heating for R$_{2}$. In order to explain the observed baseline voltage in R$_3$, we introduce a temperature dependent Seebeck coefficient and resistance. In a linear approximation the temperature dependent Seebeck coefficient is written as $S(T)= S_0 \left( 1+ \zeta\Delta T \right)$, where $\Delta T$ the local temperature increase and $\zeta = 1/T_0$ with $T_0$ the common device temperature \cite{Mott1936}. The resistivity of a metal increases with temperature as $\rho (T) = \rho_0 \left(1+\alpha \Delta T \right)$, where $\alpha$ is in the order of $10^{-3}$ K$^{-1}$ for most metals. Now, $R_{3b}$ refers to the sum of the changes in Seebeck coefficient and resistance due to Joule heating. $R_{3b}$ shows similar behavior as $R_{1b}$, since $R_{3b}$ describes essentially the temperature dependence of $R_{1b}$. Moreover, $R_{2b}$ is slightly modified by Peltier heating combined with a temperature dependent Seebeck coefficient and resistance. Nevertheless, we do not find a exponential relationship between $R_{2b}$ and $L$, indicating that the Joule heating induced Seebeck voltage is dominating.

Furthermore, we observe a spin voltage in the higher harmonic responses (shown in Fig.\ \ref{fig:data2}a and \ref{fig:data2}b). The spin signal in $R_2$ can be associated both with the temperature dependence of the spin valve signal in $R_1$ and with thermal spin injection. However, the observed signal has the opposite sign and contradicts with earlier measurements \cite{Slachter2010, Jedema_N_410_2001, Garzon2005}. The exact origin may be found in the spin-dependent Peltier effect \cite{Gravier2006} or interface scattering \cite{Hatami2009} and will not be further discussed here. The spin signal in $R_3$ is due to a change in the spin valve signal caused by Joule heating. From this measurement, we can derive the spin valve temperature dependence $\gamma$, defined as $R_{s}(T) = R_{s}(1-\gamma\Delta T)$. We found a $\gamma$ of approximately 1\%, in good agreement with earlier results \cite{Jedema_N_410_2001, Garzon2005}.

The magnitudes of the induced Seebeck voltages have been calculated with the thermoelectric model using a 3D finite element method (FEM). To obtain the linear response voltage, we have used the following parameters. The conductivities that were taken are measured separately and we found $\sigma_{\text{Py}} = 4.3 \times 10^6$~Sm$^{-1}$, $\sigma_{\text{Cu}} = 4.3 \times 10^7$~Sm$^{-1}$, and $\sigma_{\text{Au}} = 2.2 \times 10^7$~Sm$^{-1}$. For Seebeck coefficients, we took $S_{\text{Py}}=-20$~$\mu$VK$^{-1}$ \cite{Uchida2008}, $S_{\text{Cu}}=1.6$~$\mu$VK$^{-1}$, and $S_{\text{Au}}=1.7$~$\mu$VK$^{-1}$ \cite{Kittel1995}. The thermal conductivities are derived from the electrical conductivity via the Wiedeman-Franz law. In addition to thermal currents, the model describes the charge distribution in the device accurately as well. For short $L$ we find, therefore, an exponential decrease in the baseline resistance, as discussed previously. Interestingly, for larger separations due to thermoelectric effects, the baseline diminishes more moderately.  The calculated data for $R_1$ is plotted in Fig.\ \ref{fig:data2}a (blue dots) and matches the measurements remarkably well. We also incorporated the temperature dependence of the resistance and Seebeck coefficient into our thermoelectric model and the simulations are displayed in Fig.\ \ref{fig:data2}c and \ref{fig:data2}d. The slope obtained from the simulation of $R_{2b}$ deviates from the measured data by approximately a factor two. Therefore, we deduce that the Joule heating in the device is two times larger than expected. This discrepancy is ascribed to the oxidation of the Py and the interface resistance of the Au contacts, thereby reducing the thermal conduction. For the calculation of $R_{3b}$, we corrected for this, and obtained a perfect agreement between the simulation of $R_{3b}$ and the experimental data.

To confirm our analysis we excluded charge current effects completely. Therefore, we have measured a similar device with an interrupted Cu strip as shown in Fig.\ \ref{fig:sem}b. Heat conduction can still occur through the SiO$_2$, but charge transport is eliminated. We found a nonzero baseline resistance of 1.85~m$\Omega$ for $L=\text{300~nm}$, significantly smaller than without the interruption. This change is mainly due to the difference in thermal conductivity for SiO$_2$ and Cu. FEM calculations predicted for $k_{SiO_2} = 1$~Wm$^{-1}$K$^{-1}$ a resistance $R_{1b}$ of 1.9 m$\Omega$, in perfect agreement with the observed value. For $R_{2b}$ we found 3.75~$\mu V mA^{-2}$, compared to 4.4~$\mu V mA^{-2}$ for the calculations.

In conclusion, we have demonstrated that thermoelectric effects play an important role in nanoscale spin valve devices. These effects have been employed to locally raise and probe the electron temperature at the interface of two materials and the experimental results are in good agreement with basic thermoelectric rules. We found that in the analysis of the spin valve measurements, the interplay between the Peltier and Seebeck effect lead to a significant increase in baseline resistance. By probing the second and third harmonic response separately, higher order thermal effects are observed. In general, these findings open new possibilities for future caloritronic applications using localized electron temperature control.

We would like to acknowledge B. Wolfs, S. Bakker and J.G. Holstein for technical
assistance. This work is part of the research program of the Foundation for Fundamental Research on Matter (FOM) and supported by DynaMax, NanoNed and the Zernike Institute for Advanced Materials.


\end{document}